\documentclass[12pt,a4paper]{article}

\usepackage{epsfig}
\usepackage{amsmath,amsfonts,amssymb}
\usepackage{cite}
\usepackage{colortbl}
\usepackage{pifont}

\providecommand{\openone}{\leavevmode\hbox{\small1\kern-3.8pt\normalsize1}}

\newcommand{\be}{\begin{equation}}
\newcommand{\ee}{\end{equation}}

\newcommand{\afb}{A_\text{FB}}
\newcommand{\ac}{A_\text{C}}

\newcommand{\minv}{m_{t\bar{t}}}
\newcommand{\bmin}{\beta_\text{min}}

\newcommand{\bmu}{\mathcal{B}_\mu}

\newcommand{\gmu}{\mathcal{G}_\mu}

\newcommand{\omf}{\omega^4}

\newcommand{\gM}{\gamma^\mu}

\newcommand{\la}{\lambda^a}

\begin{document}

\begin{center}
\begin{Large}
{\bf Boosting the $t \bar t$ charge asymmetry}
\end{Large}

\vspace{0.5cm}
J. A. Aguilar--Saavedra$^a$, A. Juste$^{b,c}$, F. Rubbo$^c$ \\
{\it $^a$ Departamento de F\'{\i}sica Te\'orica y del Cosmos and CAFPE, \\
Universidad de Granada, E-18071 Granada, Spain \\
$^b$ Instituci\'o Catalana de Recerca i Estudis Avan\c{c}ats (ICREA), Barcelona, Spain \\
$^c$ Institut de F\'{\i}sica d'Altes Energies (IFAE), Barcelona, Spain}
\end{center}

\begin{abstract}
We propose a kinematical enhancement of the $t \bar t$ charge asymmetry at the LHC by selecting events with the $t \bar t$ centre of mass frame highly boosted along the beam axis. This kinematical selection increases the asymmetries and their significance up to a factor of two in a rather model-independent fashion. Hence, it can be a perfect complement to enhance model discrimination at the LHC.
\end{abstract}

\section{Introduction}

The observation of an unexpectedly large forward-backward (FB) asymmetry in $t \bar t$ production by the Tevatron experiments constitutes one of the most solid hints of new physics in the top sector. The latest inclusive values reported by the CDF and D0 Collaborations~\cite{Aaltonen:2011kc,Collaboration:2011rq} are around two standard deviations above the Standard Model (SM)
predictions~\cite{Kuhn:1998jr,Antunano:2007da,Bernreuther:2010ny,Ahrens:2011uf,Hollik:2011ps} and even larger departures are found for other related measurements. But the experimental situation is not yet clear, with the CDF result pointing at a strong mass dependence of the asymmetry which is not confirmed by the D0 Collaboration. On the other side, the CMS~\cite{afbCMS2} and ATLAS~\cite{afbATLAS} Collaborations have measured the charge asymmetry in $t \bar t$ production at the Large Hadron Collider (LHC),
\begin{equation}
\ac = \frac{N(\Delta > 0) - N(\Delta < 0)}{N(\Delta > 0) + N(\Delta < 0)} \,,
\label{ec:AC}
\end{equation}
with $\Delta = |\eta_t|- |\eta_{\bar t}|$ (CMS) or $\Delta = |y_t|- |y_{\bar t}|$ (ATLAS), being $\eta$, $y$ the pseudo-rapidity and rapidity, respectively, of the top (anti)quark and $N$ standing for the number of events. The small, negative asymmetries measured by both experiments are hard to reconcile with positive deviations at Tevatron~\cite{AguilarSaavedra:2011ug} but the still large errors in the present measurements allow for small positive asymmetries, compatible with a Tevatron excess. In this situation, it is of great interest to explore possible ways of enhancing the LHC charge asymmetry, in order to have an independent test of the Tevatron anomalies as sensitive as possible.

The $t \bar t$ charge and FB asymmetries only arise in the $q \bar q \to t \bar t$ subprocess, since the $gg$ initial state is symmetric. At the partonic level, the kinematics of $q \bar q \to t \bar t$ can be described by the partonic centre of mass (CM) energy $\hat s$ (which equals the $t \bar t$ invariant mass squared $m_{t \bar t}^2$) and the CM opening angle $\theta$ between the top and the initial quark. A third relevant quantity, independent of the former two, is the boost of the partonic CM with respect to the laboratory frame. This boost can conveniently be parameterised by the velocity of the $t \bar t$ system along the beam axis in the laboratory frame,
\begin{equation}
\beta = \frac{|p_{t}^z+p_{\bar t}^z|}{E_{t}+E_{ \bar t}} \,
 \end{equation}
being $p^z$, $E$ the momentum along the beam axis and energy, respectively.\footnote{Note that the velocity is related to the partonic momentum fractions $x_{1,2}$ as $\beta=|x_1-x_2|/(x_1+x_2)$.}
An asymmetry enhancement can be achieved by a phase space selection involving one or more of these three variables $\minv$, $\theta$, $\beta$, at the expense of reducing the data sample and thus the statistics. In this respect, it is important to stress here that
the $t \bar t$ invariant mass is not a suitable parameter to increase the asymmetry. For some models, like extra $Z'$~\cite{Jung:2009jz} or $W'$~\cite{Cheung:2009ch} bosons, the asymmetry grows with $\minv$ while for other models, such as light $s$-channel colour octets~\cite{Barcelo:2011vk,Tavares:2011zg,Alvarez:2011hi,AguilarSaavedra:2011ci} or scalars exchanged in the $u$ channel~\cite{Shu:2009xf} the 
$\minv$ profile of the asymmetry can be completely different and $\ac$ may even become negative at high $\minv$. Indeed, the asymmetry dependence on the $t \bar t$ invariant mass is most useful for model discrimination\cite{AguilarSaavedra:2011hz}. 

\begin{figure}[htb]
\begin{center}
\epsfig{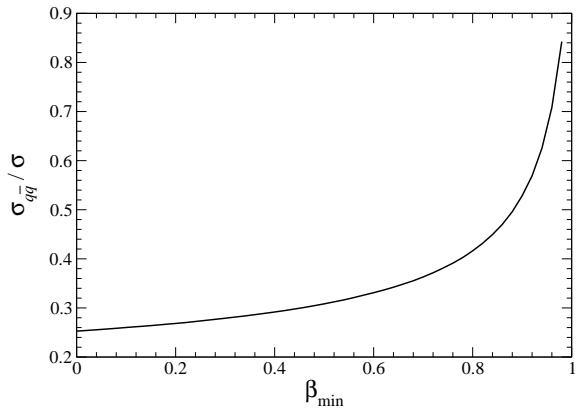} 
\caption{Relative fraction of $q \bar q \to t \bar t$ events as a function of the minimum $t \bar t$ velocity.}
\label{fig:qqfrac}
\end{center}
\end{figure}

Previous literature already includes proposals on this topic. In the so-called forward asymmetry~\cite{Hewett:2011wz}
\begin{align}
& A_\text{fwd} = \frac{N(|y_t| > y_C) - N(|y_{\bar{t}}| > y_C)}{N(|y_t| > y_C) + N_{\bar t}(|y_{\bar{t}}| > y_C)} \,,
\label{ec:AF}
\end{align}
with $y_C$ some fixed rapidity cut, a selection is effectively placed on the angle $\theta$ (also depending on $\beta$), to obtain a charge asymmetry larger than the inclusive one. Similar results are found~\cite{Arguin:2011xm} by requiring the leptonic top quark in the central detector with $|\eta| < 2.5$ and the hadronic one in the forward region $|\eta| > 2.5$ (with decay products in $|\eta| < 4.5$), a selection which also affects both $\theta$ and $\beta$. In both proposals, the largest improvement is found for SM extensions in which the asymmetry is most significant at small $\theta$, due to the exchange of a light particle (a $Z'$ or $W'$ boson) in the $t$ channel. On the other hand, for simple new physics models involving new particles in $s$ or $u$ channels these kinematical selections do not bring such an improvement~\cite{AguilarSaavedra:2011ug}, and the statistical significance of the asymmetry even decreases with respect to the inclusive measurement. (A larger asymmetry may still be an advantage if the measurement is dominated by systematic uncertainties.) In this Letter we explore an alternative way of increasing the asymmetry, by using a single cut on the $t\bar t$ velocity $\beta$ but without any restriction on $\theta$ or $\minv$. 
As it is well known, one of the reasons for the smallness of the charge asymmetry at the LHC, compared to the Tevatron, is the smaller fraction of `asymmetric' $q \bar q \to t \bar t$ events in the total $t \bar t$ sample, dominated by $gg$ fusion at LHC energies. For $t \bar t$ events originating from $q \bar q$ annihilation, the partonic CM frame tends to be more boosted along the beam axis, due to the much higher average momentum fractions for quarks than for antiquarks in $pp$ collisions. Therefore, the requirement of a minimum $t \bar t$ velocity $\bmin$ increases the $q \bar q$ fraction in the sample, as it can be seen in Fig.~\ref{fig:qqfrac}, calculated at the tree-level using CTEQ6L1~\cite{Pumplin:2002vw} parton density functions (PDFs) for a CM energy of 7 TeV.
The choice of $\beta$ instead of the momentum $|p_{t}^z+p_{\bar t}^z|$ to increase the asymmetry~\cite{Wang:2010du} is motivated by its small correlation with other energy variables such as $\minv$, as well as by the fact that it is experimentally a more robust observable, less affected by uncertainties on the jet energy scale and resolution.
Also, this simple cut on $\beta$ is an alternative to more sophisticated analyses~\cite{Bai:2011uk} to enhance the $q \bar q$ fraction by using a likelihood function built of several kinematical variables of the $t \bar t$ pair and its decay products, whose practical application may suffer from important systematic uncertainties.

\section{Asymmetries at the parton level}
\label{sec:2}

After these introductory considerations, we proceed to investigate how the asymmetry is increased in SM extensions accommodating the Tevatron measurements, and to which extent this increase is model-independent. For this, we select three simple benchmark models: (i) an axigluon $\gmu$~\cite{Ferrario:2008wm}; (ii) a $Z'$ boson; (iii) a colour-triplet scalar $\omf$, which correspond to the exchange of new particles in the $s$, $t$, $u$ channels in $q \bar q \to t \bar t$, respectively. Their quantum numbers and interactions are summarised in Table~\ref{tab:lagr}. More specifically, our benchmark models are:
\begin{table}[htb]
\begin{center}
\begin{tabular}{ccll}
Label & Spin & Rep. & \multicolumn{1}{c}{Interaction Lagrangian} \\
\hline
$\gmu$ & 1 & $(8,1)_0$
  & $- \left( g_{ij}^q \bar q_{Li} \gM \frac{\la}{2} q_{Lj} 
  + g_{ij}^u \bar u_{Ri} \gM \frac{\la}{2} u_{Rj}  + g_{ij}^d \bar d_{Ri} \gM \frac{\la}{2} d_{Rj} \right) \mathcal{G}_\mu^a$ \\[1mm]
$\bmu$ & 1 & $(1,1)_0$ 
  & $-\left( g_{ij}^q \bar q_{Li} \gM q_{Lj} 
  + g_{ij}^u \bar u_{Ri} \gM u_{Rj} + g_{ij}^d \bar d_{Ri} \gM d_{Rj} \right) \bmu $ \\[1mm]
$\omf$ & 0 & $(3,1)_{-\frac{4}{3}}$
  & $- g_{ij} \varepsilon_{abc} \bar u_{Rib} u_{Rjc}^c \, \omega^{4a\dagger} + \text{h.c.}$ \\[1mm]
\end{tabular}
\caption{Quantum numbers and relevant interactions for the new particles considered in our benchmark models.\label{tab:lagr}}
\end{center}
\end{table}
\begin{itemize}
\item Axigluon: A neutral colour-octet vector $\gmu$ with axial couplings $g_{ii}^q = - g_{ii}^u = - g_{ii}^d$, exchanged in the $s$ channel in $q \bar q \to t \bar t$. There are different proposals~\cite{Barcelo:2011vk,Tavares:2011zg,Alvarez:2011hi,AguilarSaavedra:2011ci} of light colour octets consistent with the $t \bar t$ invariant mass measurements at Tevatron and LHC; here for simplicity we consider this new particle to be heavy enough not to be produced on shell, and replace its propagator by a four-fermion interaction~\cite{delAguila:2010mx}.
\item $Z'$ boson: A neutral (colour- and isospin-singlet) vector boson $\bmu$ with flavour-violating couplings, exchanged in the $t$ channel in $u \bar u \to t \bar t$. We consider only $g_{13}^u$ non-zero (right-handed couplings) as preferred by $B$ physics constraints.
\item Colour-triplet scalar: A charge 4/3 colour-triplet $\omf$ with a flavour-violating coupling $g_{13}$, exchanged in the $u$ channel in $u \bar u \to t \bar t$. 
\end{itemize}
The parameters for these three models are chosen so as to have new physics contributions\footnote{Next-to-leading order (NLO) SM contributions are not included in our analysis; the total asymmetries when these are included as well can be approximately obtained by adding the SM NLO asymmetry to the new physics contributions presented.} to the inclusive charge asymmetry $\ac^\text{new} \simeq 0.04$.\footnote{Note that for the $Z'$ model there is a minimum positive value $\ac^\text{new} \simeq 0.04$ consistent with the total $t\bar t$ cross section at Tevatron~\cite{AguilarSaavedra:2011hz}; for a better comparison between $s$, $t$ and $u$ channels we have also chosen $\ac^\text{new} \simeq 0.04$ for the axigluon and colour-triplet scalar.} 
 For the heavy axigluon we select $C/\Lambda^2 = 1.86~\text{TeV}^{-2}$. For the $Z'$ boson we choose a  ``light'' mass $M = 150$ GeV with a coupling $g_{13}^u = 0.55$, for which the forward enhancement of the asymmetry at $\theta \sim 0$ is much more pronounced~\cite{AguilarSaavedra:2011ug} and the differences with $s$- and $u$-channel exchange larger. For the scalar we use an intermediate mass $M = 700$ GeV and a coupling $g_{13}=2.1$. The new physics contributions to the Tevatron inclusive asymmetry are $\afb^\text{new} = 0.189$ ($\gmu$), 0.194 ($Z'$), 0.190 ($\omf$). These three benchmark points are in agreement with the constraints on cross sections at Tevatron and LHC used in Refs.~\cite{AguilarSaavedra:2011ug,AguilarSaavedra:2011hz}.

\begin{figure}[htb]
\begin{center}
\begin{tabular}{ccc}
\epsfig{file=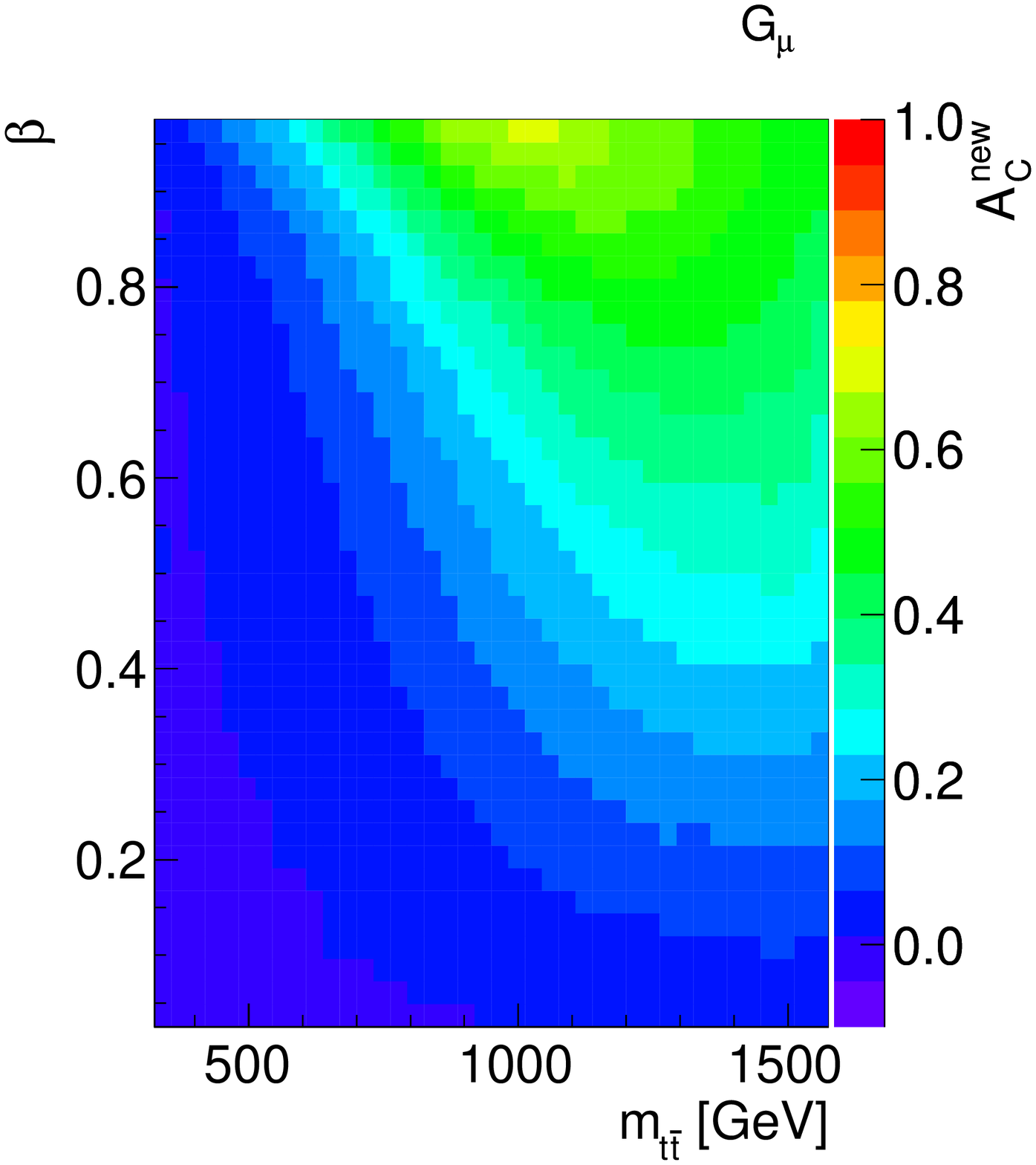,height=5cm,clip=} &
\epsfig{file=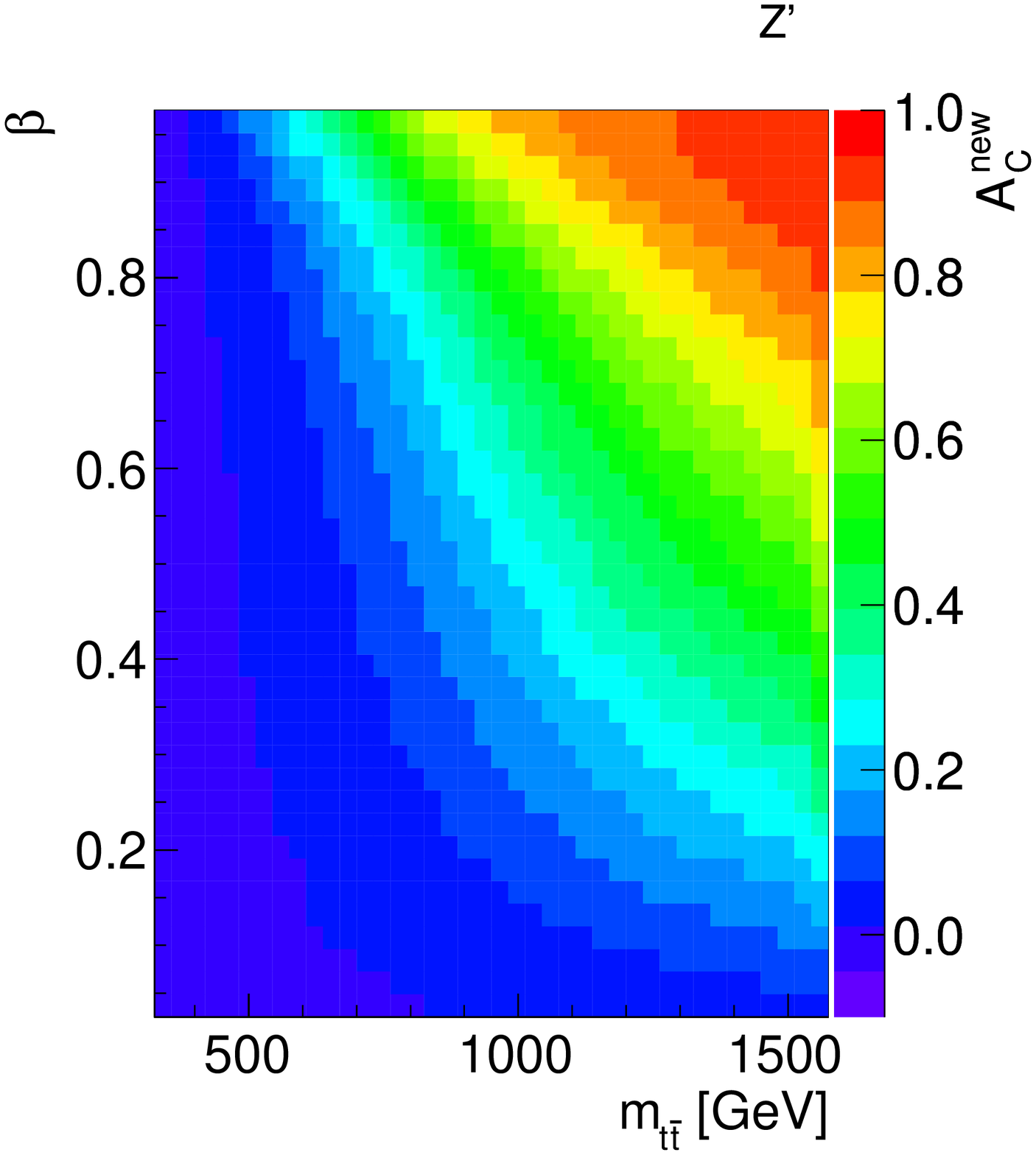,height=5cm,clip=} &
\epsfig{file=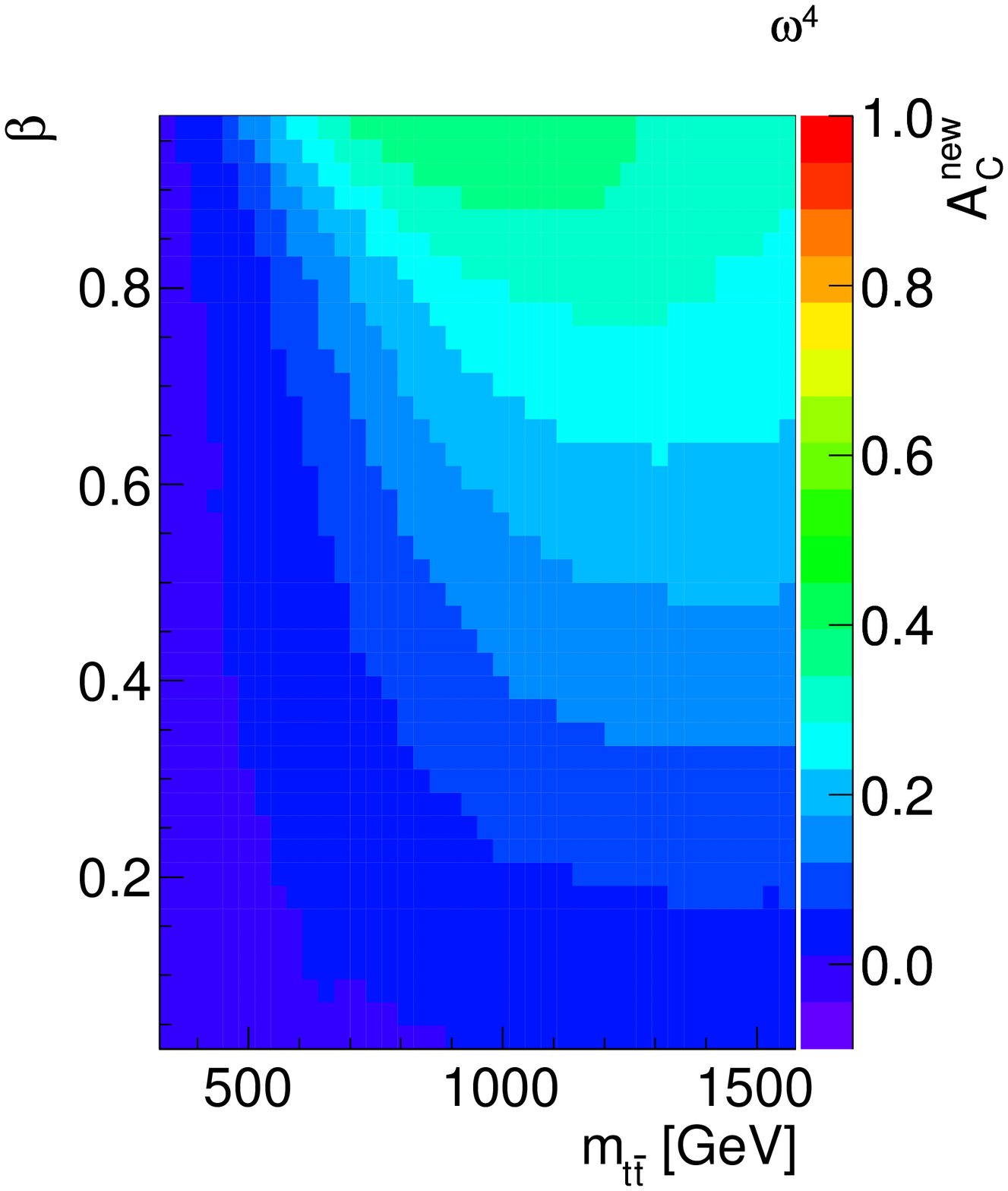,height=5cm,clip=}
\end{tabular}
\caption{Charge asymmetry as a function of the $t \bar t$ invariant mass and velocity in the laboratory frame, for the three benchmark models.}
\label{fig:A2D}
\end{center}
\end{figure}

\begin{figure}[htb]
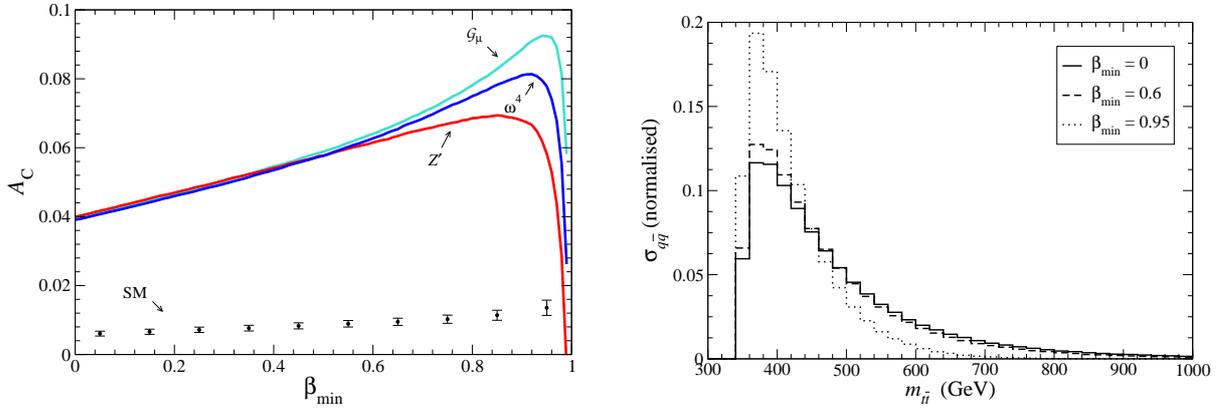

\begin{center}
\begin{tabular}{ccc}
\epsfig{file=Figs/asym-beta,width=7.5cm,clip=} & \quad &
\epsfig{file=Figs/mtt-beta,width=7.5cm,clip=}
\end{tabular}
\caption{Left: new physics contributions to the charge asymmetry as a function of the lower cut $\bmin$, for the three benchmark models (solid lines), and SM contribution (points with error bars). Right: normalised $\minv$ distribution for $q \bar q \to t \bar t$ in the SM, for several values of $\bmin$.}
\label{fig:asym-beta}
\end{center}
\end{figure}

The charge asymmetry as a function of $\minv$ and $\beta$ is presented in Fig.~\ref{fig:A2D} for the three benchmark models with the parameters above mentioned.
In all cases we observe a significant asymmetry increase with $\beta$, showing the usefulness of requiring a minimum $t \bar t$ velocity $\bmin$ to enhance it. In Fig.~\ref{fig:asym-beta} (left) we plot the actual effect of such a cut at the parton level. (The integrated asymmetries in Fig.~\ref{fig:asym-beta} are related to the differential ones in Fig.~\ref{fig:A2D} by convolution with PDFs and integration over $\beta > \bmin$ and all the $\minv$ range.) 
We observe that for the three models the integrated asymmetries increase monotonically up to $\bmin \sim 0.6$
in a model-independent fashion, as it is expected from the kinematical enhancement of the $q \bar q$ fraction in the sample. This feature is quite desirable, since it allows to use a cut on $\beta$ to enhance the asymmetry while retaining $\minv$ as a very useful variable for model discrimination. The small SM contribution, calculated with {\tt MC@NLO}~\cite{Frixione:2002ik}, is also displayed with error bars corresponding to the Monte Carlo statistical uncertainty. It exhibits the same relative increase with respect to the inclusive value, as expected. (The total asymmetry is the sum of SM and new physics contributions to a good approximation, so one can safely focus on new physics contributions and add the SM contribution at the end if desired.)
For larger $\bmin$ some differences between the models begin to show up, 
originated by the different $\minv$ dependence of the asymmetry in each case (see Fig.~\ref{fig:A2D}), and the fact that $t \bar t$ events with higher longitudinal boost tend to have a smaller invariant mass, due to the strong suppression of the PDFs at high momentum fraction. This correlation is clearly observed in Fig.~\ref{fig:asym-beta} (right), where we plot the normalised $t \bar t$ invariant mass distribution for $q \bar q \to t \bar t$ in the SM, for $\bmin = 0,\;0.6,\;0.95$. For a moderate value $\bmin \sim 0.6$ the normalised $\minv$ distributions are hardly affected by the cut, ensuring that the asymmetry enhancement is model-independent. Nevertheless, this is no longer the case for much larger values such as $\bmin = 0.95$. This lower average $\minv$ at high $\bmin$ is precisely the origin of the sudden drop of the asymmetries for $\bmin \gtrsim 0.95$, despite the larger $q \bar q$ fraction, see Fig.~\ref{fig:qqfrac}. 

\begin{figure}[htb]
\begin{center}
\epsfig{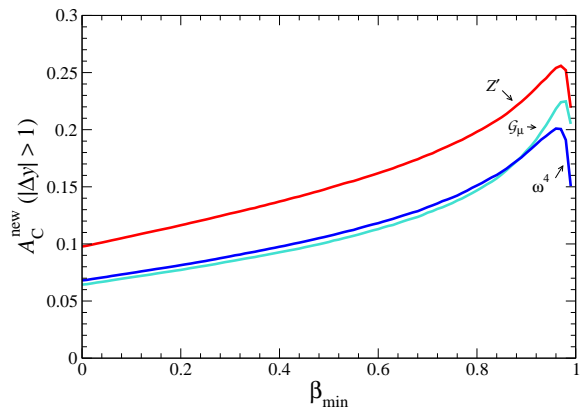}
\caption{Charge asymmetry at high rapidities $|\Delta y|>1$ as a function of the lower cut $\bmin$, for the three benchmark models.}
\label{fig:asym1-beta}
\end{center}
\end{figure}

Finally, it is worth pointing out that the asymmetry increase with a cut on $\beta$ is complementary to other possible model-dependent enhancements, for example restricting the range of variation of $\theta$. To illustrate this, we show in Fig.~\ref{fig:asym1-beta} the charge asymmetry as a function of $\bmin$ for the same models displayed in Fig.~\ref{fig:asym-beta} but after the requirement $|\Delta y|>1$, a cut which places a selection on the angle $\theta$.\footnote{Since $\Delta y$ is invariant under boosts along the beam axis, $|\Delta y| = \left|\log \frac{1+\beta_t \cos \theta}{1-\beta_t \cos \theta}\right|$, being $\beta_t=\sqrt{1-4 m_t^2/\hat s}$ the velocity of the (anti)top quark in the CM frame.}
There are two remarkable features apparent from this plot. First, the cut $|\Delta y|>1$ increases the asymmetry to a larger extent for the $Z'$ model. As we have mentioned in the introduction, this is expected~\cite{AguilarSaavedra:2011ug} since the forward enhancement is much more pronounced in models with $t$-channel exchange of light particles. Second, the asymmetries increase in nearly the same fashion up to $\bmin \sim 0.6$, in agreement with the results shown in Fig.~\ref{fig:asym-beta}.

\section{Asymmetries at the reconstruction level}
\label{sec:3}

Having established the enhancement of the asymmetry for $t \bar t$ events boosted along the beam axis,
it is necessary to investigate further whether this selection may really constitute an advantage in a real experiment or the potential increase is washed out by detector and reconstruction effects. For this purpose, we have performed a fast simulation of three event samples for the axigluon model, with $C/\Lambda^2 = 0.93$, $1.86$, $2.94~\text{TeV}^{-2}$, resulting in $\ac^\text{new} = 0.02$, $0.04$, $0.06$.
The selection of these three benchmarks with different values of $\ac^\text{new}$ is intended to explore the sensitivity increase depending on the actual value of the asymmetry.
The events are generated with {\tt Protos}~\cite{AguilarSaavedra:2008gt} and include the top quark and $W$ boson decay with spin effects. Parton showering and hadronisation is performed by {\tt Pythia}~\cite{Sjostrand:2006za} and the package {\tt AcerDet}~\cite{RichterWas:2002ch} is used to perform a fast detector simulation and reconstruction of objects such as charged leptons and jets. We focus on the semileptonic $t \bar t$ decay channel, selecting events which fulfill the following criteria:
\begin{itemize}
\item exactly one lepton (electron or muon) with transverse momentum $p_T>20$~GeV and pseudorapidity $|\eta|<2.5$;
\item missing transverse energy $E_T^{miss}>25$~GeV;
\item at least four jets with $p_T>25$~GeV and $|\eta|<4.5$ and at least one $b$-tagged jet.
\end{itemize}
In particular, extending the jet acceptance to $|\eta|<4.5$ (though $b$ tagging is only available for $|\eta|<2.5$) is important to maintain a good acceptance for boosted events~\cite{Arguin:2011xm,Kagan:2011yx}.
We assume a per-jet $b$ tagging efficiency of $60\%$ for jets originating from a $b$ quark with $|\eta|<2.5$, and a total efficiency for lepton triggering and identification of 70\%.
This event selection is similar
to those used in the recent measurements by the ATLAS and CMS collaborations~\cite{afbATLAS, afbCMS2} and has an efficiency of
$\sim 25\%$ for semileptonic $t\bar{t}$ events. In order to compute the asymmetry, we perform a simplified $t\bar{t}$ event reconstruction by looping over neutrino solutions and jet permutations, and selecting the configuration that minimises the function
\begin{equation}
\chi^2 = \frac{(m_{j_1j_2}-M_W)^2}{\sigma_W^2} + \frac{(m_{j_1j_2j_3}-m_t)^2}{\sigma_t^2} + \frac{(m_{\ell \nu j_4}-m_t)^2}{\sigma_t^2} \,,
\end{equation}
where $m_{j_1j_2}$ ($m_{j_1j_2j_3}$) is the reconstructed invariant mass of the $W$ boson (top quark) candidate decaying hadronically, $m_{\ell \nu j_4}$ is the invariant mass of the top quark decaying leptonically, and we take $M_W=80.4$~GeV, $m_t=172.5$~GeV, $\sigma_W=10$~GeV, and $\sigma_t=20$~GeV. The chosen values for $\sigma_W$ and $\sigma_t$ are representative of the $W$ boson and top mass reconstruction resolutions 
provided by the fast simulation.
The neutrino transverse momentum is set equal to the vector $E_T^{miss}$ and the $z$ component of its momentum is obtained by solving the quadratic equation $(p_\ell+p_\nu)^2=M_W^2$. In case two real solutions exist, both of them are considered in the $\chi^2$ minimisation over configurations. If no real solution exists,  the neutrino pseudo-rapidity is set to be equal to the one of the charged lepton. Only the leading four jets in $p_T$ are considered as candidates for the $b$ quarks.  All selected jets are considered as candidates for the
hadronic $W$ boson decay, skipping the $b$-tagged jets whenever there are at least two jets that are not $b$ tagged. The configuration yielding the lowest $\chi^2$ is used to reconstruct the top and anti-top quark four-momenta. The charge asymmetry is computed using $\Delta = |y_t|-|y_{\bar t}|$. The fraction of events with the sign of $\Delta$ correctly reconstructed is $\sim 70\%$, a performance comparable to that achieved in fully simulated events by the ATLAS and CMS 
collaborations~\cite{afbATLAS, afbCMS2}. At $\bmin \simeq 0.9$ some efficiency is lost for this event selection primarily because of the detector acceptance cut of $|\eta|<  2.5$ for charged leptons
and $b$ tagging.

\begin{figure}[htb]
\begin{center}
\begin{tabular}{ccc}
\epsfig{file=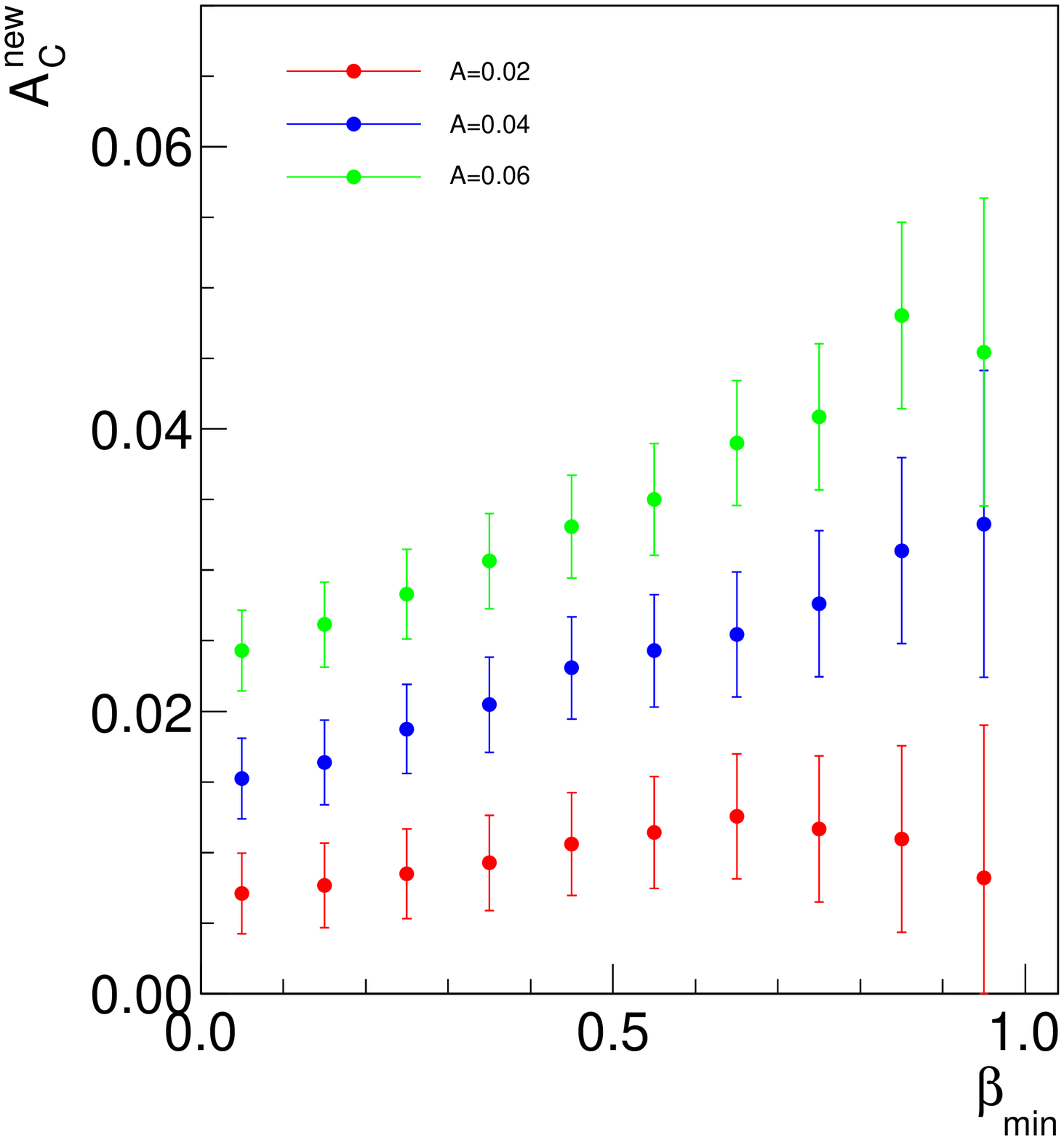,height=7.5cm,clip=} & \quad &
\epsfig{file=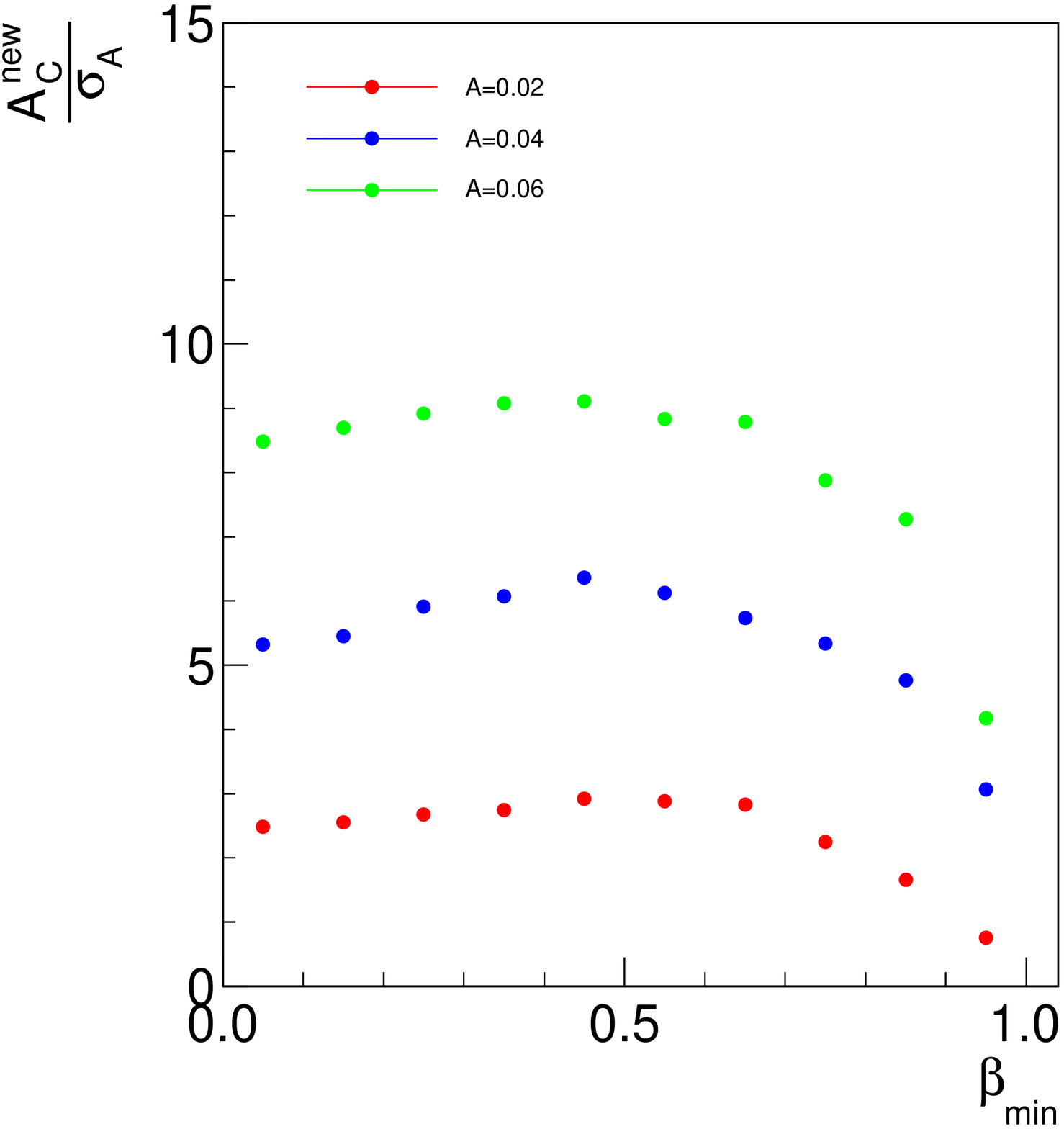,height=7.5cm,clip=} \\[2mm]
\epsfig{file=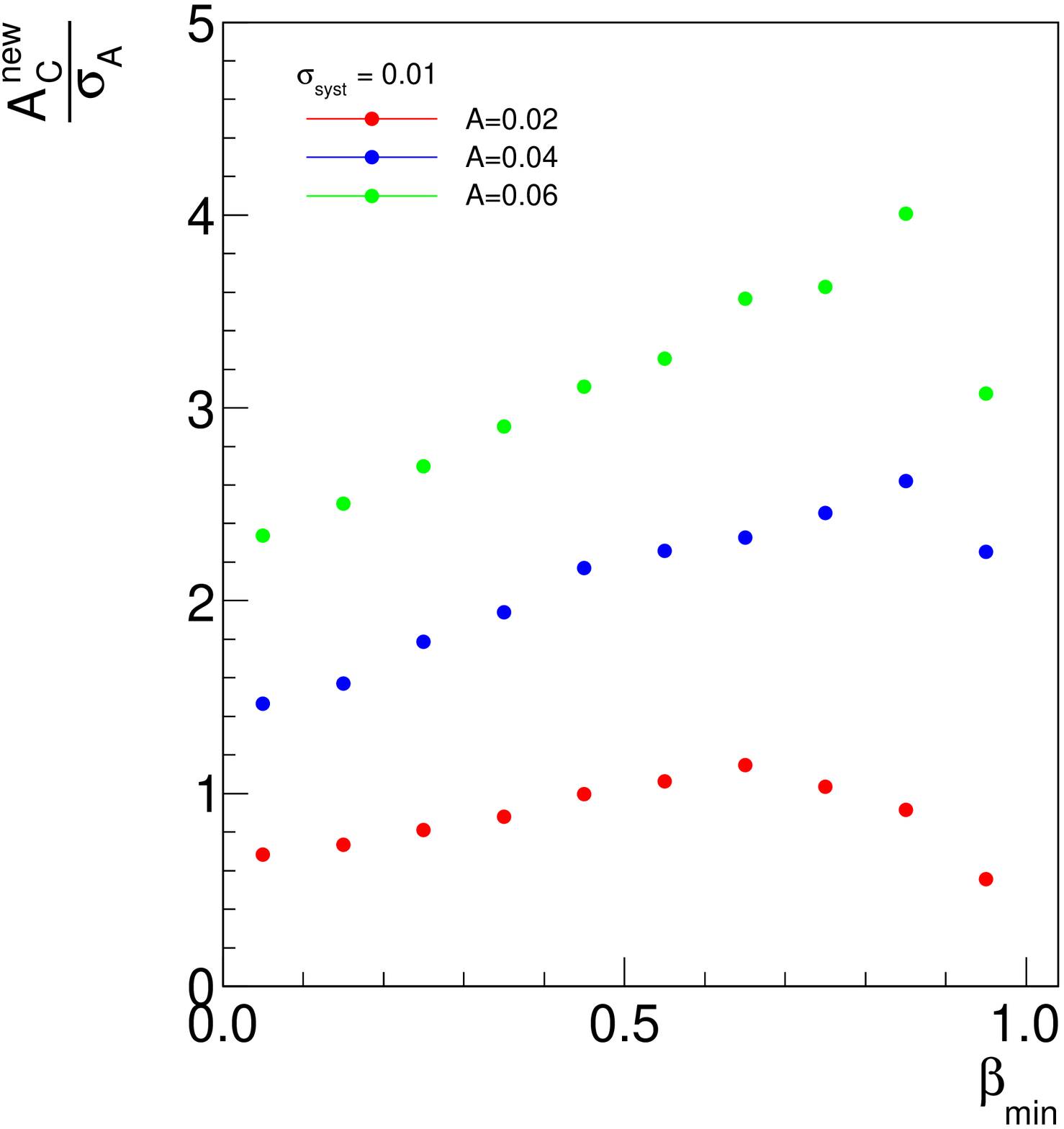,height=7.5cm,clip=} & &
\epsfig{file=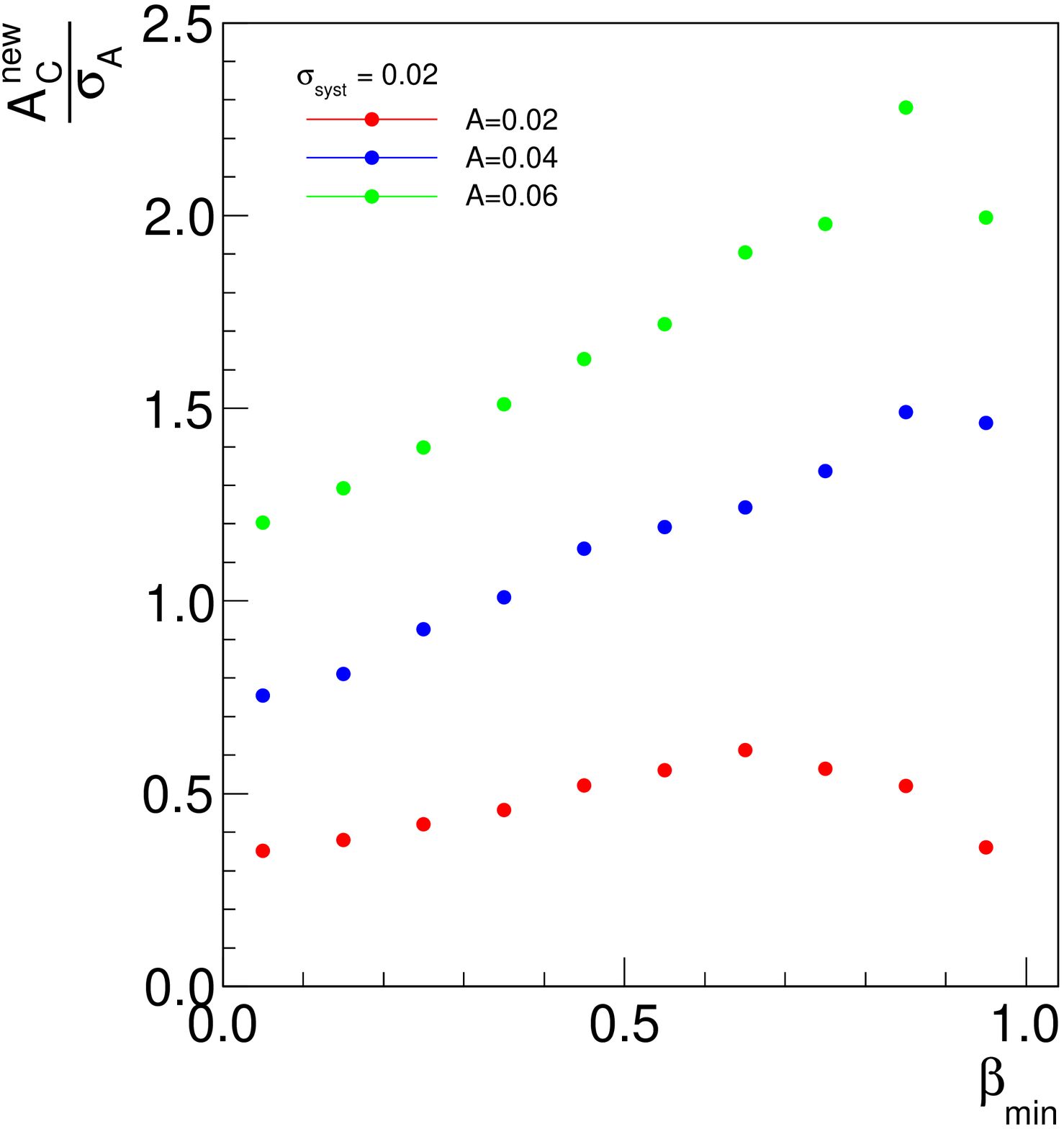,height=7.5cm,clip=}
\end{tabular}
\caption{Top, left: charge asymmetry at the reconstruction level, as a function of $\bmin$ (only statistical uncertainties, corresponding to a luminosity of 10 fb$^{-1}$, are shown). Top, right: statistical significance of the asymmetry. Bottom: significance assuming systematic uncertainties of 0.01 (left), 0.02 (right).}
\label{fig:asym-reco}
\end{center}
\end{figure}

\begin{figure}[htb]
\begin{center}
\begin{tabular}{ccc}
\epsfig{file=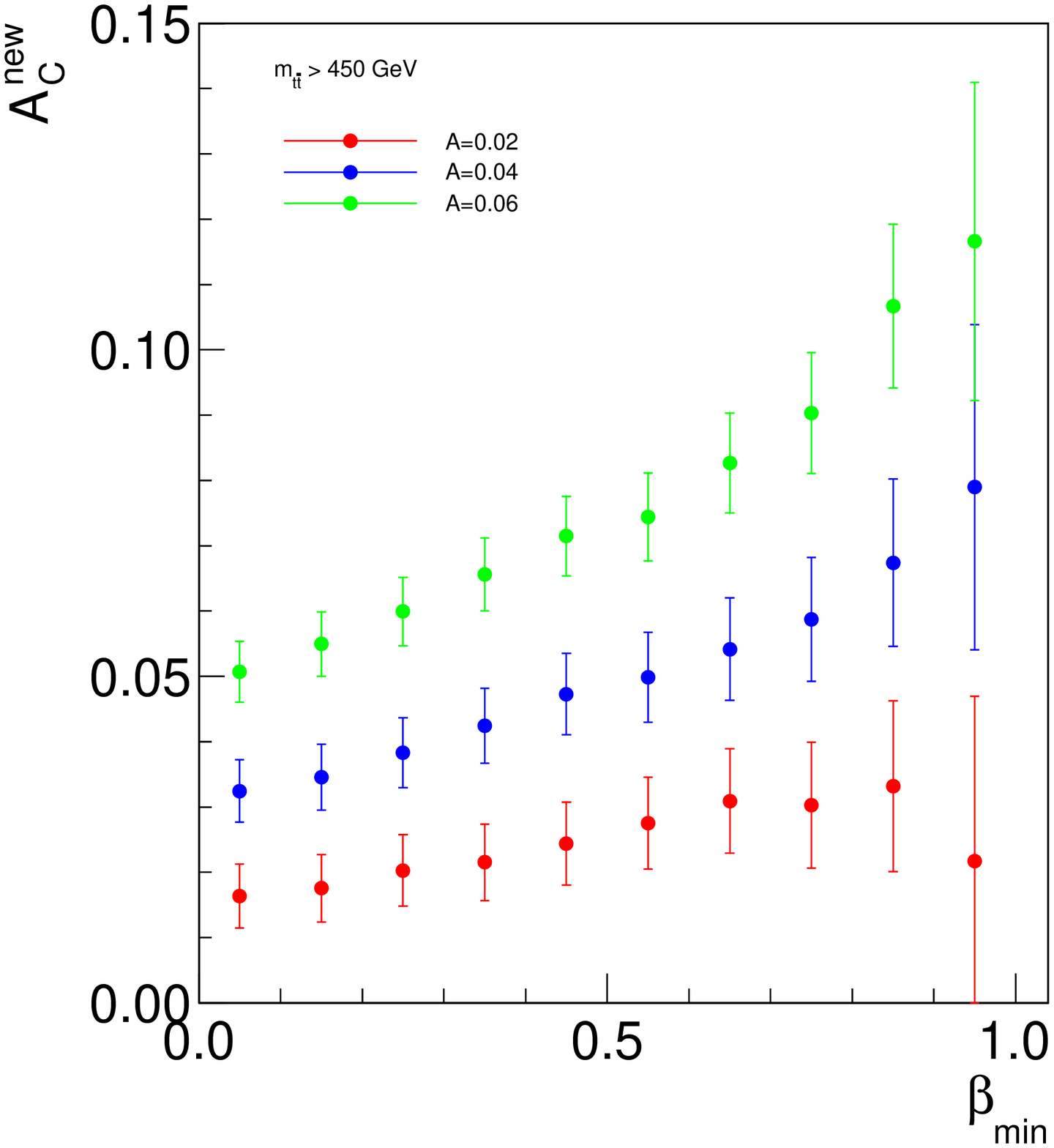,height=7.5cm,clip=} & \quad &
\epsfig{file=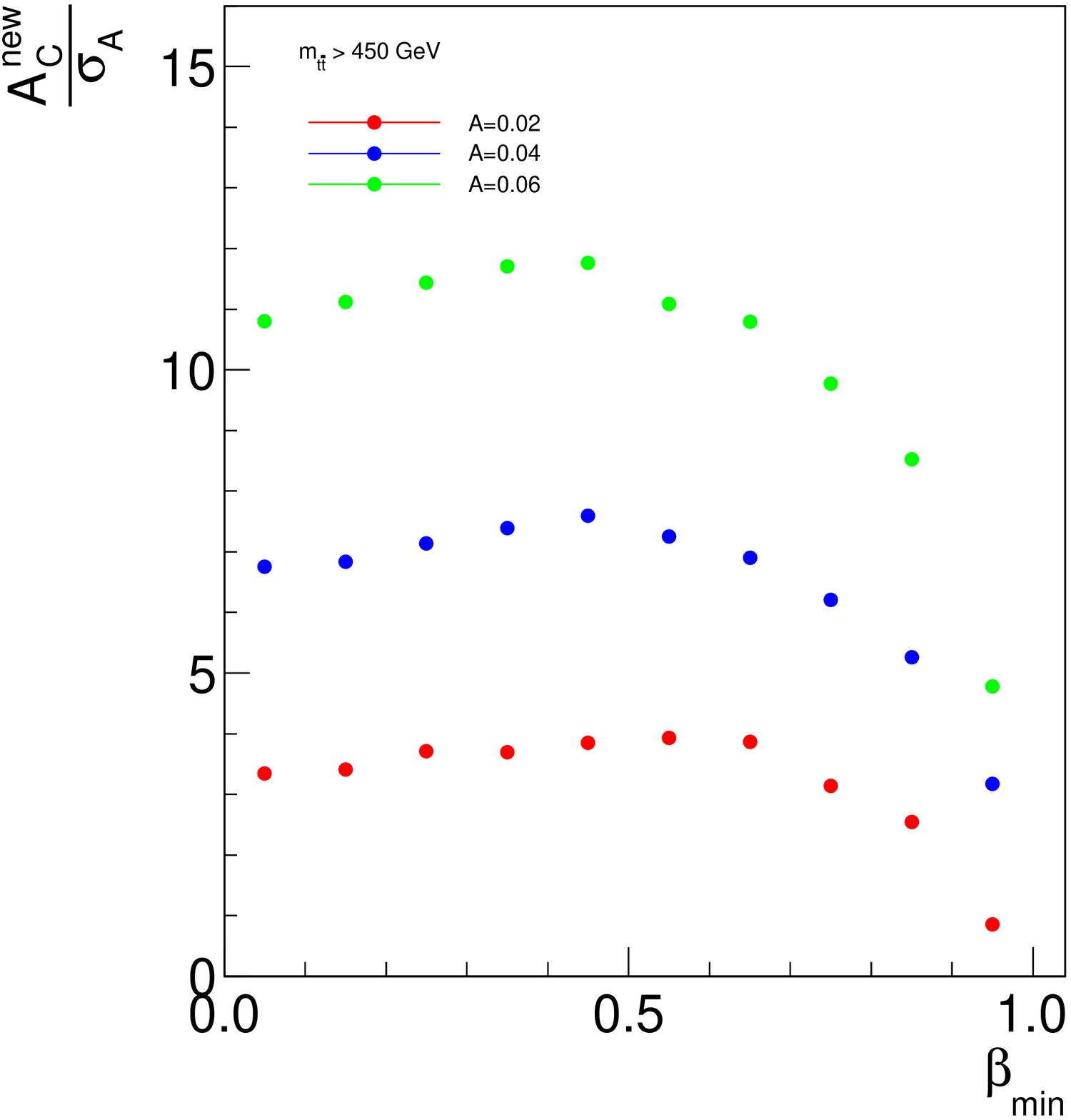,height=7.5cm,clip=} \\[2mm]
\epsfig{file=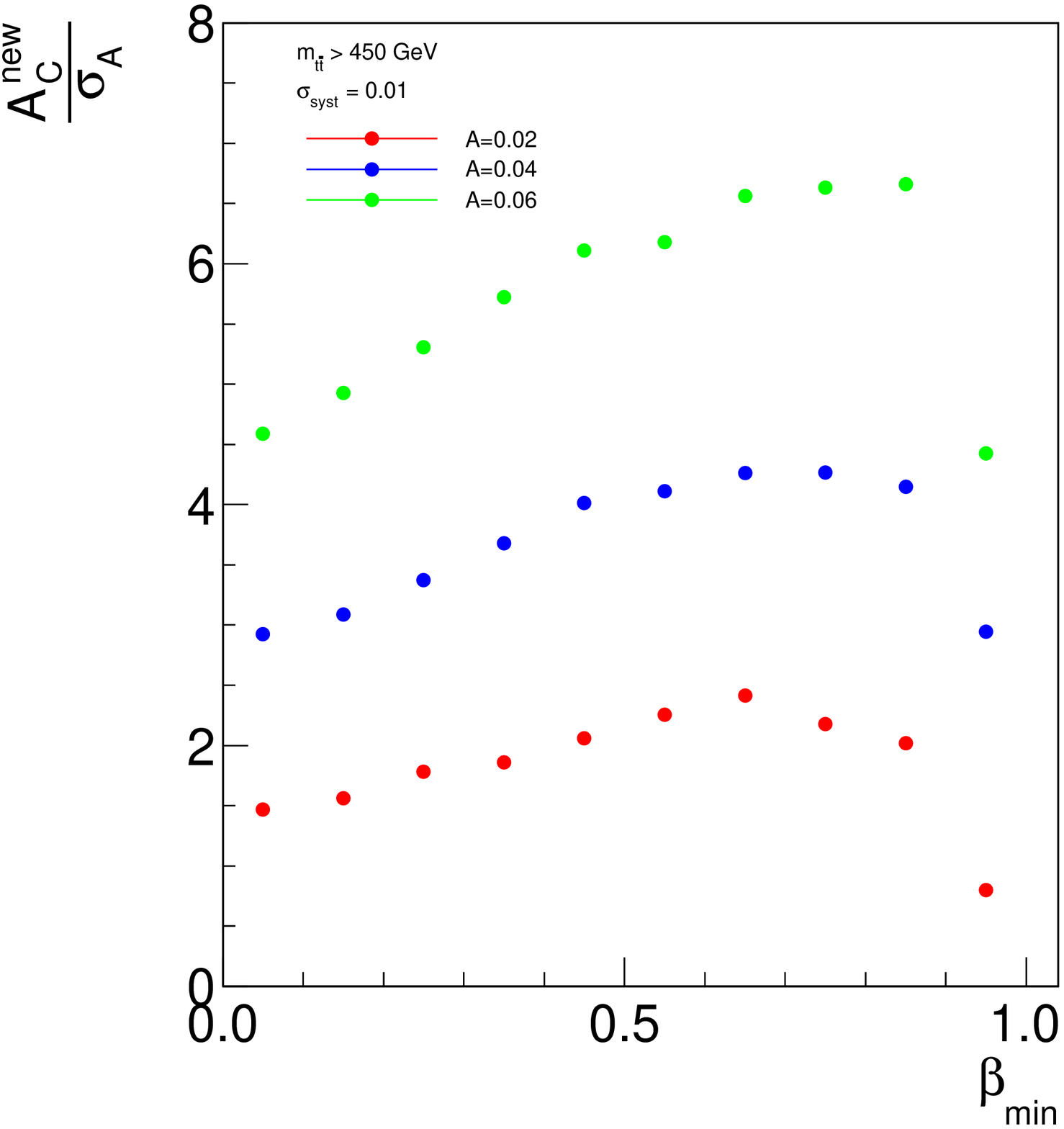,height=7.5cm,clip=} & &
\epsfig{file=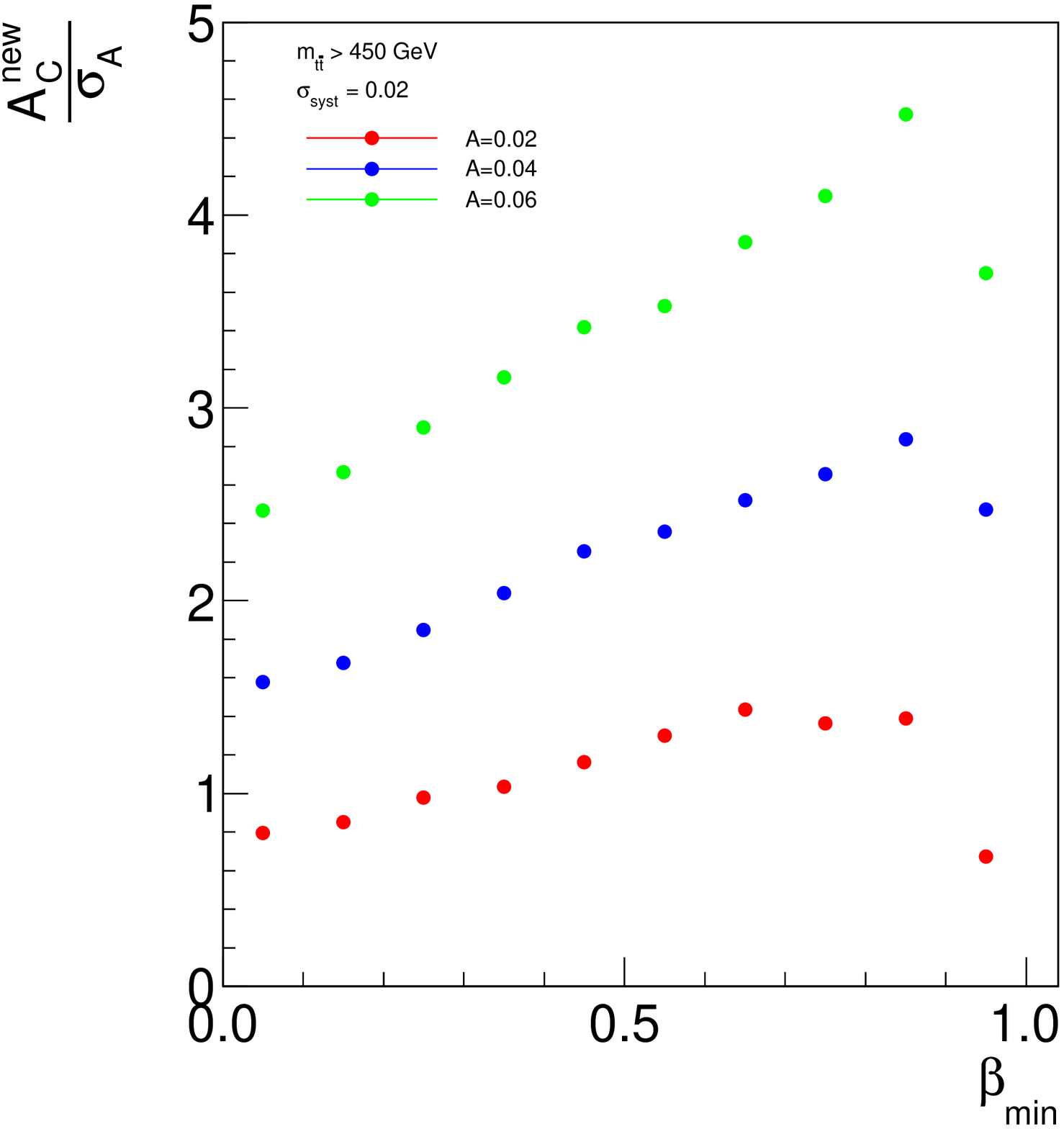,height=7.5cm,clip=}
\end{tabular}
\caption{The same as in Fig.~\ref{fig:asym-reco} but for $\minv > 450$ GeV.}
\label{fig:asym450-reco}
\end{center}
\end{figure}

We do not attempt here an unfolding of the simulated measurements to reconstruct the parton-level quantities, as this requires a very delicate analysis. Instead, we present our results at the reconstruction level and we do not include backgrounds. The latter simplification is justified by the relatively small background fraction ($\sim 20\%$) found with this kind of event selection in the experimental analyses~\cite{afbCMS2,afbATLAS} and the fact that, after subtraction from the data, the background is found to contribute in a small way to both the statistical and systematic uncertainty of
the measurements.
Figure~\ref{fig:asym-reco} (top, left) shows the asymmetry for the three heavy axigluon benchmarks as a function of $\bmin$, in bins of 0.1 (only statistical uncertainties are shown). The upper right panel corresponds to the statistical significance of the asymmetries $A/\sigma_A$, assuming a luminosity of 10 fb$^{-1}$.
We can observe that a cut on $\beta$ already leads to some statistical improvement at the $10-20\%$ level, which is not always achieved for $s$-channel models with other proposals~\cite{Hewett:2011wz,Arguin:2011xm}. Nevertheless, the real advantage of having larger asymmetries results when systematic uncertainties are taken into account, which eventually dominate for large data samples.
The lower two plots in Fig.~\ref{fig:asym-reco} show the significance assuming common systematic uncertainties of 0.01 (left) and 0.02 (right), independent of $\bmin$ in a first approximation. These assumed values represent reasonable extrapolations of the total systematic uncertainty ($\sim 0.025$) in the existing LHC experimental results~\cite{afbCMS2,afbATLAS}, which is dominated by uncertainties in the physics modeling
of $t \bar t$ production. (A careful assessment of systematic uncertainties at higher values of $\bmin$, such as {\em e.g.} 
those resulting from increased jet energy scale uncertainties for forward jets, is detector-dependent
and beyond the scope of this study.)
A lower cut on $\beta$ can increase the significance up to a factor of $1.6-2$, depending on the value of the asymmetry and the size of the systematic uncertainties. Larger enhancements in the overall significance are possible in a more model-dependent way through more stringent cuts on $\beta$. Besides, in the benchmarks considered in this section the asymmetry also grows with the $t \bar t$ invariant mass (see the next section for illustration of other possibilities). Then, it is interesting to check that at higher invariant masses a cut on $\beta$ still improves the significance in this case. This is shown in Fig.~\ref{fig:asym450-reco}, where we see that a cut on $\beta$ increases the significance of the asymmetry, up to factors of $1.5-2$ which depend on the size of systematic uncertainties.

\section{Improving model discrimination}
\label{sec:4}

\begin{figure}[htb]
\begin{center}
\epsfig{file=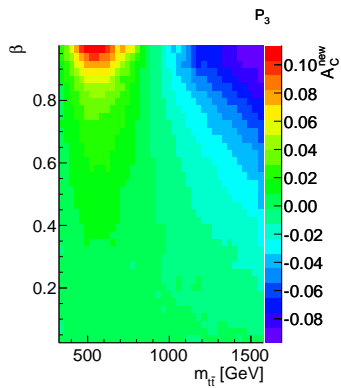,height=5cm,clip=}
\caption{Charge asymmetry as a function of the $t \bar t$ invariant mass and velocity in the laboratory frame, for model $\text{P}_3$ (see the text).}
\label{fig:A2Dl}
\end{center}
\end{figure}

\begin{figure}[htb]
\begin{center}
\begin{tabular}{ccc}
\epsfig{file=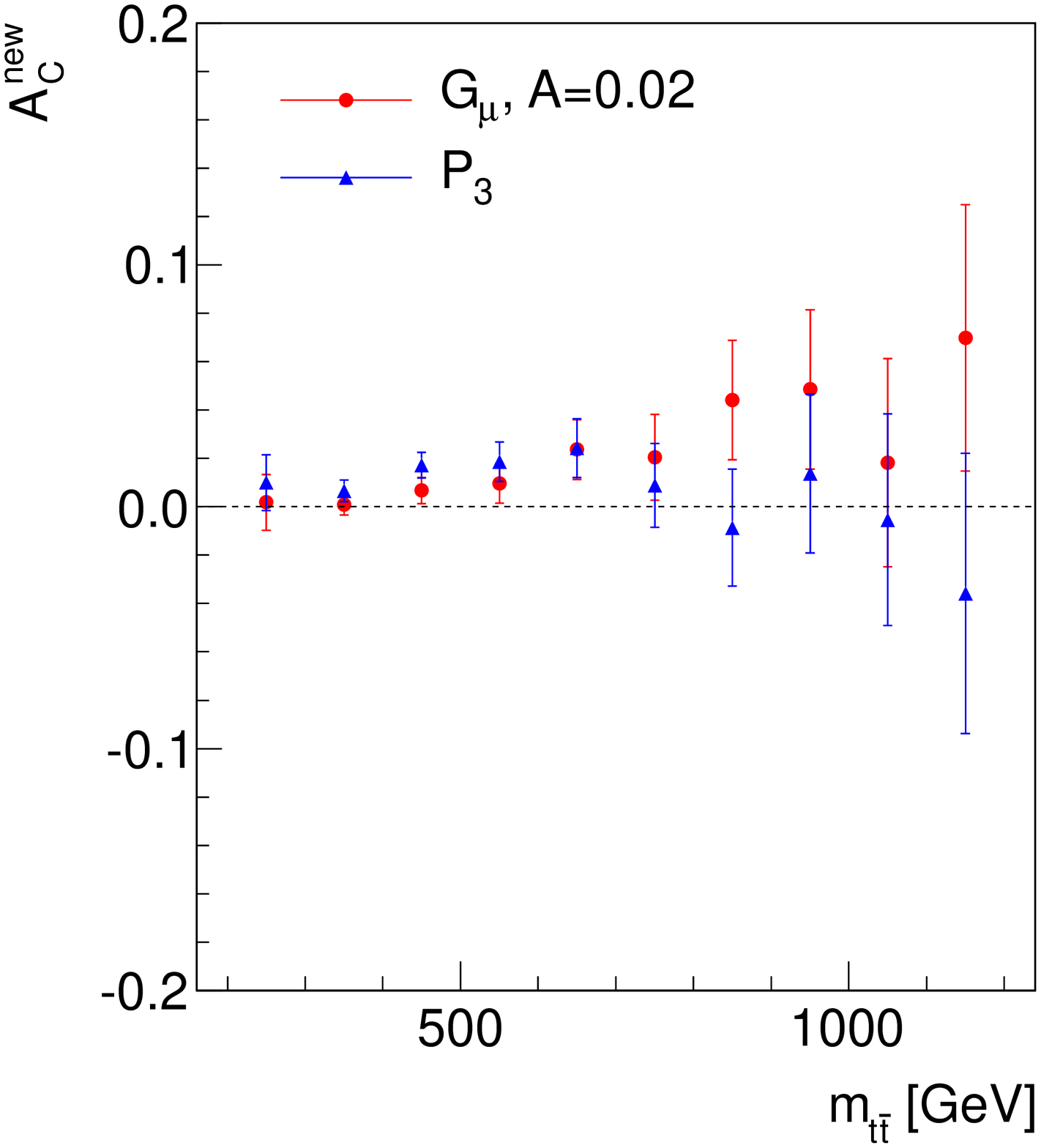,width=7cm,clip=} & \quad \quad &
\epsfig{file=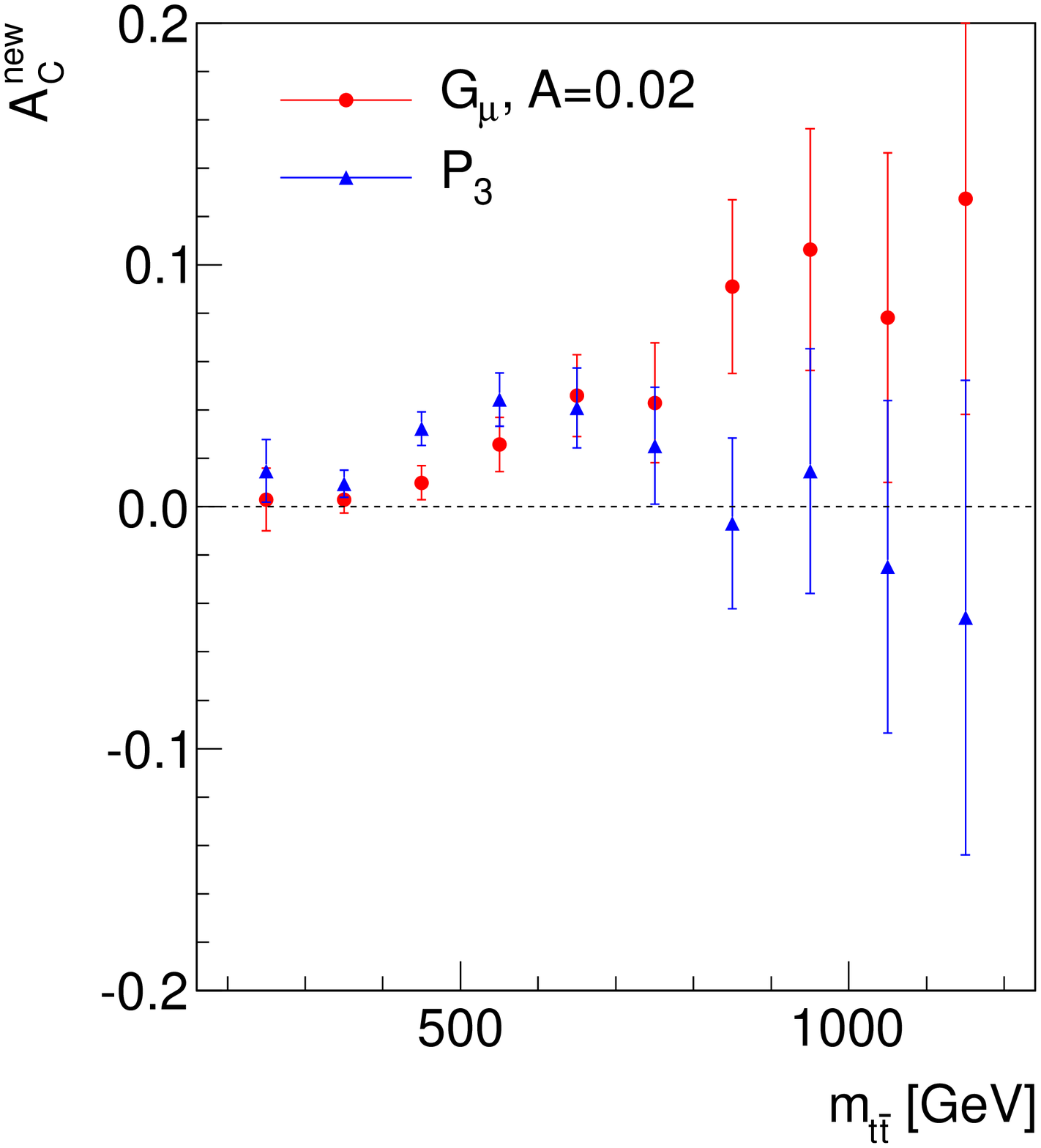,width=7cm,clip=}
\end{tabular}
\caption{Charge asymmetry at the reconstruction level, without cut (left) and for $\bmin = 0.6$ (right), for two colour octet models (see the text). Only statistical uncertainties, corresponding to a luminosity of 10 fb$^{-1}$, are shown.}
\label{fig:asym-reco2}
\end{center}
\end{figure}

We finally illustrate how our proposal for a kinematical enhancement of the charge asymmetry constitutes a perfect complement to the model discrimination by the analysis of the $\minv$ dependence of the asymmetry~\cite{AguilarSaavedra:2011ci,AguilarSaavedra:2011hz}. We have selected two difficult scenarios for the LHC with a small charge asymmetry, consistent with the most recent CMS and ATLAS measurements. The first one is the heavy axigluon $\gmu$ of the previous section with $\ac^\text{new} = 0.02$. The second one is model $\text{P}_3$ in Ref.~\cite{AguilarSaavedra:2011ci}, a colour octet with a mass $M = 870$ GeV and a large width $\Gamma = 0.6 M$, yielding $\ac^\text{new} = 0.016$. This latter model has the particularity that the asymmetry becomes negative above the resonance threshold (see Fig.~\ref{fig:A2Dl}) and this effect is testable at the LHC.
We present in Fig.~\ref{fig:asym-reco2} the (differential) charge asymmetry as a function of $\minv$ after simulation and reconstruction, without any cut on $\beta$ (left) and setting $\bmin = 0.6$ (right).
The error bars represent the statistical uncertainty in each bin. Both plots are nearly identical except for a scale factor: the asymmetries are roughly a factor of two larger when $\bmin = 0.6$ is required. This confirms again our result that the asymmetry enhancement is model-independent for moderate cuts on the $t \bar t$ velocity $\beta$. Of course,
the increase of the asymmetries makes model discrimination easier once systematic uncertainties, not included in these plots, are taken into account.

\section{Summary}

In this Letter we have proposed a kinematical enhancement of the charge asymmetry in $t \bar t$ production at the LHC, by using the velocity $\beta$ of the $t \bar t$ CM in the laboratory frame. 
Being an adimensional quantity (in natural units), $\beta$ is expected to be less sensitive to experimental uncertainties associated to the jet energy scale and resolution. In contrast with other proposals, which require a different event selection or a different definition of the asymmetry, a lower cut $\beta \geq \bmin$ is easy to implement in the current ATLAS and CMS analyses to increase the asymmetry and its significance. This asymmetry increase is independent, and complementary, to other model-dependent enhancements such as a lower cut on $|\Delta y|$. For moderate values $\bmin \lesssim 0.6$ the asymmetry enhancement is found to be model-independent. Therefore, this kinematical selection of events with larger asymmetry is also a perfect complement to an analysis of the $\minv$ dependence of the asymmetry for the purpose of model discrimination.

\section*{Acknowledgements}

We thank M. P\'erez-Victoria and J. Santiago for discussions. This work has been partially supported by projects FPA2009-07496 and FPA2010-17915 (MICINN), FQM 101 and FQM 437 (Junta de Andaluc\'{\i}a) and CERN/FP/116397/2010 (FCT).

\end{document}